\documentstyle[aps,12pt,epsfig]{revtex}

\parskip=\medskipamount

\def\be{\begin{equation}}
\def\ee{\end{equation}}
\def\disp{\displaystyle}

\def\R{{\sf I\kern-.15em R}}
\def\C{\kern.1em{\raise.47ex\hbox{$\scriptscriptstyle |$}}
             \kern-.40em{\sf C}}
\def\Z{{\sf Z\kern-.45em Z}}

\begin{document}

%\begin{center}

\title
{\Large Multifractality in uniform hyperbolic lattices and in
quasi--classical Liouville field theory}
%\bigskip

\author
{Alain Comtet$^{1}$, Sergei Nechaev$^{1,2}$ and Rapha\"el
Voituriez$^{1}$}

\address
{\it $^{1}$Laboratoire de Physique Th\'eorique et Mod\`eles
Statistiques, Universit\'e Paris Sud, \\ 91405 Orsay Cedex, France}

\address
{\it $^{2}$ L D Landau Institute for Theoretical Physics, 117940,
Moscow, Russia}

\maketitle

%\end{center}

\begin{abstract}
We introduce a deterministic model defined on a two dimensional hyperbolic
lattice. This model provides an example of a non random system whose
multifractal behaviour has a number theoretic origin. We determine the multifractal exponents, discuss the termination of multifractality and
conjecture the geometric origin of the multifractal behavior in Liouville
quasi--classical field theory.
\end{abstract}

\bigskip

\section{Introduction}
\label{sect1}

The concept of multifractality consists in a scale dependence of critical
exponents \cite{man}. It has been widely discussed in the literature in the
context of various problems such as, for example, statistics of strange sets
\cite{sinai,frisch,procacc,gut}, diffusion limited aggregation \cite{dla},
wavelet transforms \cite{holsch}, conformal invariance \cite{dupl}. This
concept also proves to be useful in the context of disordered systems
\cite{douss,falko}. It was recently found that the ground state wave function
of two dimensional Dirac fermions in a random magnetic field has a multifractal
behavior. The field theoretic  investigation of the multifractality has been
undertaken in the papers \cite{kogan}, while different
interpretations of these field theoretic results from a geometrical and
physical points of view were presented in \cite{chamon} and \cite{cast}
correspondingly. This problem was recently reanalyzed in the more general
setting of systems caracterized by logarithmic correlations \cite{douss}.

Our work is mainly inspired by the approach developed in \cite{chamon} where
the authors obtain the multifractal exponents of the critical wave function by
a mapping on the problem of directed polymers
on a Cayley tree. However our starting point is different and we treat a
deterministic model defined on a Cayley tree. We take advantage of the fact
that the Cayley tree can be isometrically embedded in a space of constant
negative curvature. We assume that each vertex of the tree carries a Boltzmann
weight that depends on the hyperbolic distance from a given root point. The
corresponding partition function is a sum over a finite number of tree vertices
and has the form of a truncated Poincar\'e series. Its scaling dependence on
the size of the system is controlled by the probability distribution of traces
of $2\times2$ matrices which belong to a discrete subgroup of $PSL(2,\R)$. This
distribution, obtained by using the central limit theorem for Markov
multiplicative processes \cite{boug}, allows us to compute the multifractal
exponents and discuss the termination of multifractality. The study of the
convergence of the measure on the boundary reveals some interesting links with work of Gutzwiller and Mandelbrot \cite{gut} on multifractal measures.
Another interesting, although more speculative, aspect is connected with
a geometric approach to Liouville field theory arising in the study of low
dimensional disordered systems \cite{kogan,chamon,falko,monthus}. We suggest
that in two dimensions  our model exhibits a new type of multifractal
behavior which has a purely geometric origin.

This  paper is organized as follows. In  section \ref{sect2} we introduce the
geometrical model possessing the multifractal behavior, develop methods for
its investigation and explicitly show the number theoretic origin of
multifractality; section \ref{sect3} is devoted to applications of these
results to quasi--classical 2D Liouville field theory (LFT); the
conclusion presents some speculations regarding the applicability of our
geometric considerations to some other disordered physical systems.

\section{The model}
\label{sect2}

\begin{figure}[ht]
\begin{center}
\epsfig{file=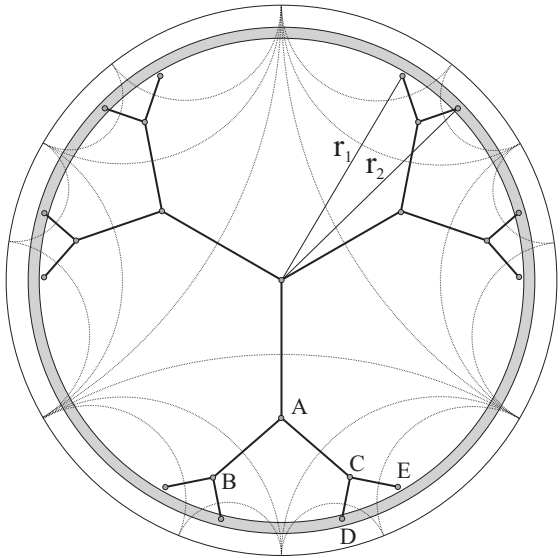,width=5cm}

\end{center}
\caption{A Cayley tree in the Poincar\'e disc. Sample points:
$A\left(-\frac{i}{2}\right)$, $\{B,C\}\left(\mp\frac{\sqrt{3}}{7}-
\frac{5i}{7}\right)$, $D\left(\frac{\sqrt{3}}{8}-\frac{7i}{8}\right)$,
$E\left(\frac{\sqrt{3}}{4}-\frac{3i}{4}\right)$.}
\label{fig:1}
\end{figure}

We begin with the investigation of geometrical properties of lattices uniformly
embedded in the hyperbolic 2-space. Lattices under consideration are defined as
follows: we construct the set of all possible orbits of a given root point
under the action of a discrete subgroup of $PSL(2,\R)$ (group of motion of the
hyperbolic 2-space). We restrict ourselves with the simplest example of
3--branching Bethe lattice (Cayley tree) which is  generated by reflections of
zero--angled curvilinear triangle---see fig.\ref{fig:1}.

The graph connecting the centers of the neighboring triangles forms a Cayley
tree isometrically embedded in the Poincar\'e unit disc (a Riemann surface of
constant negative curvature).

Consider the $n^{\rm th}$ generation of the vertices of the 3--branching Cayley tree.
Denote by $r_j(n)$ the Euclidean distance of the vertex $j$ (which
belongs to the $n^{\rm th}$ generation of the Cayley tree) from the center of the unit
disc ($1\le j\le 3\times 2^{n-1}$). The corresponding hyperbolic (geodesic)
distance $d_j(n)$ is given by:
\be \label{eq:hyp}
d_j(n)=\ln \frac{1+r_j(n)}{1-r_j(n)} \qquad
\left(r_j(n)=\tanh \frac{d_j(n)}{2}\right)
\ee
Define the generating function ${\cal Z}(q,N)$
\be \label{eq:sum}
{\cal Z}(q,N)=\sum_{n=1}^{N}
\left(\sum_{j=1}^{3\times 2^{n-1}} e^{-q\,d_j(n)}\right)
\ee
In a physical context ${\cal Z}(q,N)$ may be interpreted as a
partition function on the hyperbolic lattice with an action linear
in the length of trajectory.
%So, it is naturally to interpret such
%action as a fermionic one \cite{polyakov}.

The Bethe lattice involved can be constructed by the action of the discrete
group $\Gamma_{\theta}$ which operates on the unit disc by a set of
fractional--linear transformations. Despite the simple structure of the group
it is believed that the techniques involved are quite general and could be
easily generalized in order to cover more sophisticated lattices.

We are interested in the scaling dependence of the partition function  ${\cal
Z}(q,N)$ as a  function of the size $N$ of the system. Scaling considerations
suggest the following behaviour
\be \label{eq:scal1}
{\cal Z}(q,N)={\cal L}_{\rm max}^{-\tau^{\star}(q)}
\ee
where in our case ${\cal L}_{\rm max}=3(2^{N}-1)$ is the total number of
Cayley tree vertices in the bulk restricted by the generation $n=N$.

We show below that the critical exponent $\tau(q)$ defined as follows
\be \label{eq:scal2}
-\lim_{N\to\infty}\frac{\ln {\cal Z}(q,N)}{N}=\tau(q)
\ee
depends nonlinearly on $q$ i.e. exhibits the multicritical behavior (note that
$\tau^{\star}=\tau/\ln2$). Note that the free energy normalized per volume of the
system $f(q,N)= -\frac{\ln {\cal Z}(q,N)}{N}$ coincides with the multifractal
exponent $\tau(q)$:
\be \label{eq:scal3}
\lim_{N\to \infty}f(q,N)=\tau(q)
\ee

\subsection{Numerical results}

We first compute numerically the histogram, which counts the number of vertices
belonging to generation $n$ (properly normalized), $W_n(d)$, lying in the shell
$\left[d,\; d + \delta d\right]$.
%$\left[d_{\rm min}(n)+\delta d,\; d_{\rm min}(n) + 2\;\delta
%d\right]$, ..., $\left[d_{\rm min}(n)+ m-1\; \delta d, \;
%d_{\rm max}(N)\right]$;  $m$ is a number of "strips" in the ring between
%shortest $d_{\rm min}$ and longest $d_{\rm max}= d(N)$
%geodesics for all $n$ up to $N$.

\begin{figure}[ht]
\begin{center}
\epsfig{file=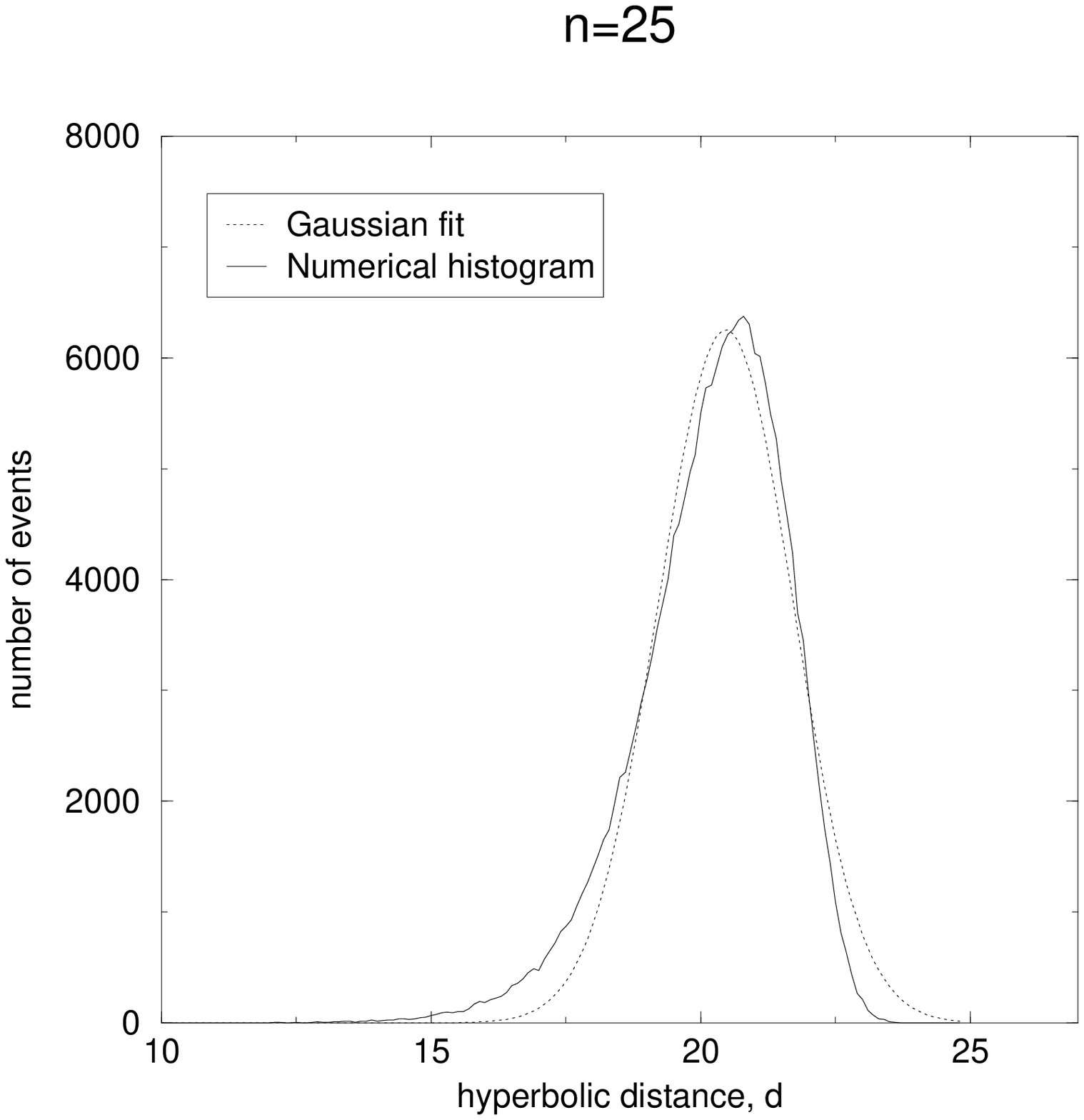,width=8cm}\epsfig{file=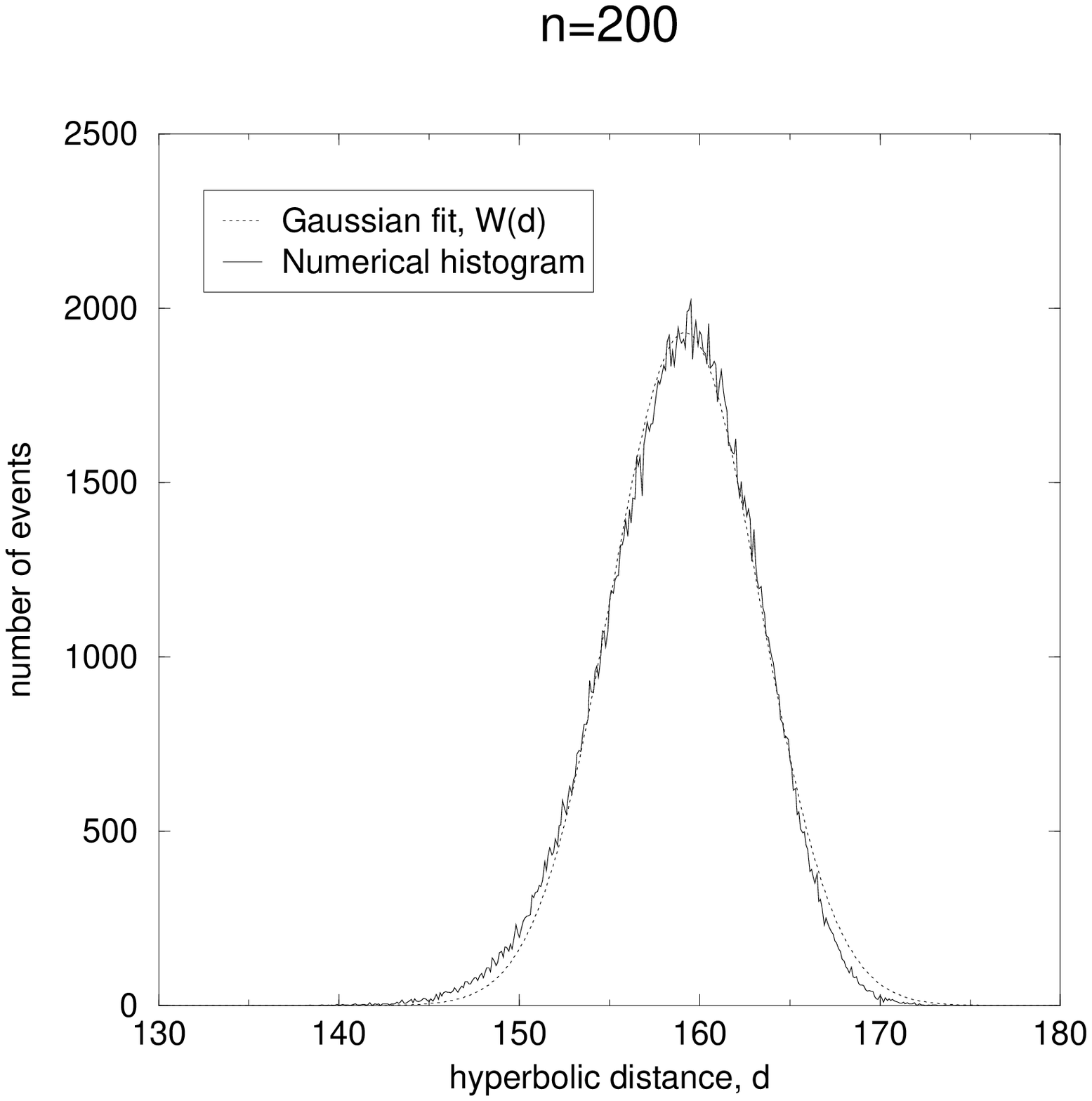,width=8.25cm}
\end{center}

%\centerline{\hspace{7cm}\epsfig{file=lat_f2a.eps,width=8cm}
%\hspace{-8.5cm} \epsfig{file=lat_f2b.eps,width=8.25cm}} \medskip

\caption{Distributions $W_n(d)$ up to normalization, compared to their
Gaussian fits: one can notice the slow convergence from strongly nonsymetric
regime for $n=25$ (a) to a Gaussian regime for $n=200$ predicted by the
central limit theorem (b).}
\label{fig:2}
\end{figure}

In our particular computations we restrict ourselves with two cases depending
on the length of the trajectories:
\begin{enumerate}
\item {\bf Short trajectories.} We enumerate all trajectories and the
computations have been carried out for all $n\in [1,N]$ up to $N=25$
generations. The figure fig.\ref{fig:2}a shows the histogram for the
distribution of hyperbolic distances for $n=25$. The absolute value of
number of events in the fig.\ref{fig:2}a depends on the particular choice of
the width of the shell $\delta$.
%For each
%generation $n$ most of geodesics lie in the "window" $\left[\left<d(n)
%\right>-\frac{\delta d}{2}, \left<d(n)\right> +\frac{\delta d}{2}
%\right]$ near $\left< d\right>$---see fig.\ref{fig:2} and the total number
%of windows is $m=20$, so $\delta  d=\frac{1}{20}( d_{\rm max}- d_{\rm
%min})$.
% The  integral histogram for all generations up to $N=18$ is compared to
%the distribution for particular (last) generation $n=18$.
It can be seen from fig.\ref{fig:2}a that the corresponding plot is highly
nonsymmetric with respect to the mean value $\left<d\right>$.
% The best numerical fit is .... \\
\item {\bf Long trajectories.} For $n=200$ the enumeration of all different
paths is very time consuming, therefore we compute numerically the histogram
$W_n(d)$ developing partial ensemble of $200\,000$ directed random walks of
$n=200$ step each. As $n\to \infty$ the distribution function
$W_n(d)$ becomes more and more symmetric in accordance with the statement that
there exists a central limit theorem for such random walks on noncommutative
groups (see the discussion below). The results of corresponding numerical
computations are presented in Fig.\ref{fig:2}b. The distribution $W_n(d)$ is
well fitted by a Gaussian function:
$$
W_n(d)=A_0\,e^{-\frac{(d-\left<d\right>)^2}{2\Delta^2}}
$$
where for $n=200$ one has: $A_0\approx 1929.96$ and depends on
normalization; $\left<d\right>\approx 159.18$; $\Delta^2\approx 17.01$.
\end{enumerate}

In spite of the fact that convergence to the Gaussian distribution is
slow, the linear dependence in $n$ of the mean value $\left<d\right>=
\gamma n$ and the variance $\left<(d-\left<d\right>)^2\right>\equiv
\Delta^2=\sigma^2 n$ is numerically evident, which permits one to get an
accurate estimate of $\gamma$ and $\sigma^2$.

\begin{figure}[ht]
\begin{center}
\epsfig{file=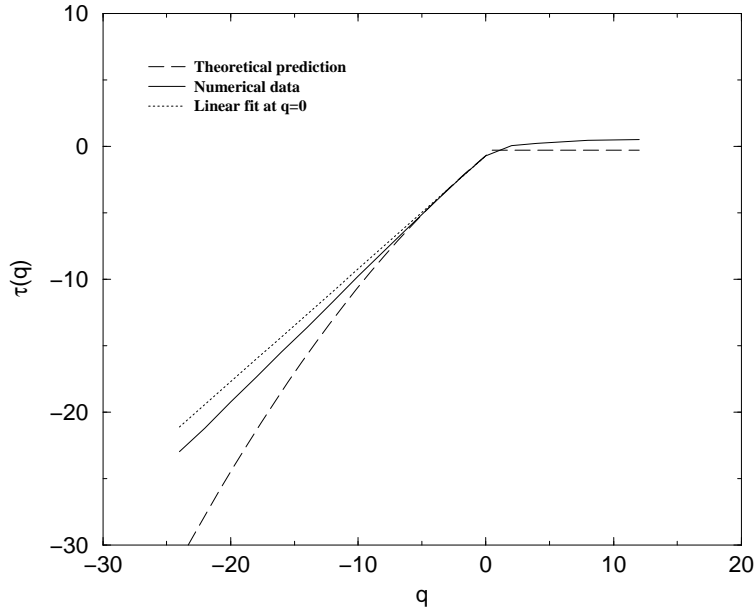,width=10cm}
\end{center}
\caption{Multicritical behavior of the exponent $\tau(q)$ for $N=40$, compared to theoretical prediction.}
\label{fig:3}
\end{figure}

The numerical computation of the probability distribution $W_n(d)$ allows one
to compute the multifractal exponent $\tau(q)$ following the definitions
(\ref{eq:scal1})--(\ref{eq:scal3}). The  corresponding results are shown in
fig.\ref{fig:3},  for $N=40$. Due to the slow convergence of the distribution, the discrepancy between numerical data (technically limited to $N\le40$) and the theoretical prediction can not be quantitatively taken into account. We here insist on the multifractal behaviour, shown by the non-linear depence on $q$.

\subsection{Analytic results}

\begin{figure}[ht]
\begin{center}
\epsfig{file=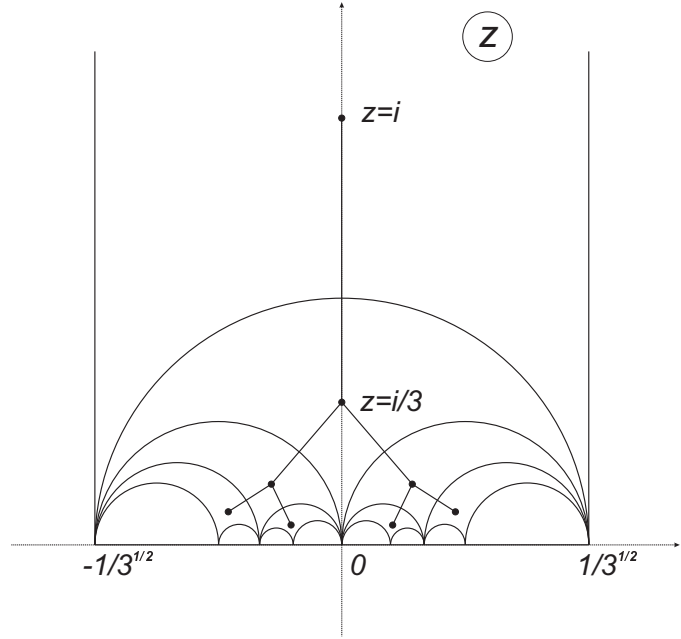,width=6cm}
\end{center}

\caption{Poincar\'e hyperbolic upper half--plane ${\cal H}^2$}
\label{fig:4}
\end{figure}

Let us return to the definition of the model and recall that the group
$\Gamma_{\theta}$ acts in the hyperbolic Poincar\'e upper half--plane ${\cal
H}^2=\{z\in\C,\,{\rm Im}(z)>0\}$ by fractional--linear
transforms\footnote{It is convenient first to define the representation of
the group $\Gamma_{\theta}$ in the Poincar\'e upper half--plane and then use
the conformal transform to the unit disc.}. The matrix representation of the
generators of the group $\Gamma_{\theta}$ is well known (see, for example
\cite{terras}), however for our purposes it is more  convenient to take a framework that consists of the composition of standard fractional--linear
transform and complex conjugacy. Namely, denoting by $\bar z$ the complex
conjugate of $z$, we consider the following action
\be
\left(\begin{array}{cc} a & b \\ c & d \end{array} \right):\
z\to\frac{a\bar z+b}{c\bar z+d}
\ee
A possible set of generators is then:
\be \label{hk}
h_0=\left(\begin{array}{cc} 1 & -2/\sqrt{3} \\ 0 & -1 \end{array} \right),\;
h_1=\left(\begin{array}{cc} 1 & 2/\sqrt{3} \\ 0 & -1 \end{array} \right),\;
h_2=\left(\begin{array}{cc} 0 & 1 /\sqrt{3} \\ \sqrt{3} & 0 \end{array} \right)
\end{equation}
Choosing the point $(x_0,iy_0)=(0,i)$ as the tree root---see fig.\ref{fig:4},
any vertex on the tree is associated with an element $\disp M_n=\prod_{k=1}^n
h_{\alpha_k}$ where $\alpha_k\in\{0,1,2\}$ and is parametrized by its complex
coordinates $z_n=M_{n}\big((-1)^n i\big)$ in the hyperbolic plane.
Strictly speaking ${\cal H}^2$ should be identified with $SL(2,\R)/SO(2)$; we
here identify an element with its class of equivalence of $SO(2)$. If one
denotes by $d(M_n)\equiv d(i,z_n)$ the hyperbolic distance between $i$ and
$z_n$, the following identity holds
\be \label{eq:cosh}
2\cosh\Big(d(M_n)\Big)={\rm Tr}(M_{n}M_{n}^{\dag})
\ee
where dagger denotes transposition.

\subsubsection{Distribution function, invariant measure on the boundary and
Lyapunov exponents}

We are interested in the distribution function $W_n(d)$ which is the
probability to find the tree vertices in generation $n$ at the distance $d$
from the root point. It means that we are looking for the distribution of the
traces for matrices $M_n$ which are the irreducible products of $n$ generators.
If we denote by $l(M_n)$ the irreducible length of the word represented by the
matrix $M_n$, then $M_n$ is irreducible if and only if $l(M_n)=n$. Such word
enumeration problem is simple in case of the group $\Gamma_{\theta}$, because
of its free product structure: $\Gamma_{\theta}\sim\Z_2\otimes\Z_2\otimes\Z_2$.
Indeed, if $\disp M_n= \prod_{k=1}^n h_{\alpha_k}$ one has $l(M_n)=n$ if and
only if  $h_{\alpha_k}\not=h_{\alpha_{k-1}}\; \forall k$. Hence we have to
study  the behavior of the random matrix $M_n$, generated by the following
Markovian process
\be \label{mar}
M_{n+1}=M_n h_{\alpha_{n+1}}\ {\rm with}\ \alpha_{n+1}=\left\{\begin{array}{ll}
(\alpha_{n}+1)\,{\rm mod}\,3 & \ \mbox{with probability $\frac{1}{2}$} 
\medskip \\
(\alpha_{n}+2)\,{\rm mod}\,3 & \  \mbox{with probability $\frac{1}{2}$} 
\end{array} \right.
\ee

We use the standard methods of random matrices and consider the entries of the
$2\times  2$--matrix $M_n$ as a 4--vector ${\cal V}_n$. The transformation
$M_{n+1}= M_n h_{\alpha}$ reads
\be
{\cal V}_{n+1}= \left(\begin{array}{ll} h_{\alpha}^{\dag} & 0 \\ 0 &
h_{\alpha}^{\dag} \end{array}\right)\; {\cal V}_{n}
\ee
This block--diagonal form allows one to restrict ourselves to the study of one
of two 2--vectors, composing ${\cal V}_{n}$, say $v_n$. Parametrizing
$v_n=(\varrho_n\cos\theta_n,\varrho_n\sin\theta_n)$ and using the relation
$d(M_n)\equiv  d_n\simeq 2\ln\varrho_n$ valid for $n\gg 1$, one gets a
recursion relation $v_{n+1}=h_{\alpha}^{\dag}v_n$ in terms of hyperbolic
distance $d_n$:
\be \label{eq:dist}
\label{dist}
\disp  d_{n+1}= d_n+\ln\,
\left[\frac{5}{3}+\frac{4}{3}\cos(2\theta_n+\varphi_{\alpha})\right]
\ee
where $\varphi_{\alpha}$ depends on the transform $h_{\alpha}$ through $\disp
\varphi_{\alpha}=(2\alpha-1)\pi/3\; (\alpha=0,1,2)$, while for the angles one
gets straightforwardly
\be \label{ang}
\tan\theta_{n+1}=h_{\tilde{\alpha}}\Big(\tan(\theta_n)\Big)
\ee
and the change $\alpha\rightarrow\tilde{\alpha}$ means the substitution
$(0,1,2)\rightarrow(1,0,2)$. Action of $h_{\alpha}$ is still
fractional--linear.

Define now three invariant measures $\mu_{\alpha}(\theta)$ corresponding to
transformations of $M_n$ ($n\gg 1$) whose last step is given by a matrix
$h_{\alpha}$. The form of (\ref{ang}) suggests to consider  the corresponding 
$\mu_{\alpha}(x)$ with $x=\tan\theta$. We are then led to study the action of
$\Gamma_{\theta}$ restricted on the real line parametrized by $x$. Interesting
properties of the average $\overline\mu=(\mu_0+\mu_1+\mu_2)/3$ have been
discussed by Gutzwiller and Mandelbrot \cite{gut}. In particular they pointed
out the connexion with the arithmetic function $\beta(\xi)$ which maps some
number $\xi\in[0,1]$ written as a continued fraction expansion
$$
\frac{1}{n_1+\disp\frac{1}{n_2+\ldots}}
$$
to the real number $\beta$ whose binary expansion is made by the sequence of
$n_1-1$ times 0, followed by $n_2$ times 1, then $n_3$ times 0, and so on. To
account for this fact, one has first to notice that the construction
(\ref{mar}) of any word $M$ in  $\Gamma_{\theta}$ is exactly encoded by the
binary representation of a real $\xi$, the $n$'s letter of this
expansion being $\alpha_{n+1}-\alpha_n+1\, {\rm mod}\,3$. The second argument,
due to Series \cite{series}, is that the real part of the vertex $M(i)$ is
precisely  the continued fraction $\mu(\xi)$. Therefore $\beta(\xi)$ has to
be proportional to the ``number'' of vertices lying in the interval $[0,\xi]$,
that is to $\overline\mu([0,\xi])$. Taking the limit $n\rightarrow\infty$
is not well defined. An alternative, which was used in this work, is to
define $\mu_{\alpha}(x)$ as the limit of the following recurrency:
\be \label{rec}
\disp \mu_{\alpha}^{(n+1)}(x)=\frac{1}{2}\left|\frac{dh_{\alpha}(x)}{dx}\right|
\sum_{\alpha'\not=\alpha}\mu_{\alpha'}^{(n)}\Big(h_{\alpha'}(x)\Big)
\ee
The symetry  of such expression leads, after summing over ${\alpha}$, to the
following relation admitting as fixed point $\overline\mu(x)$ at $n\to\infty$:
\be \label{rec2}
\mu^{(n+1)}(x)=\frac{1}{3}
\sum_{\alpha=0}^2\mu^{(n)}\Big(h_{\alpha}(x)\Big)
\left|\frac{dh_{\alpha}(x)}{dx}\right|
\ee

\begin{figure}[ht]
\begin{center}
\epsfig{file=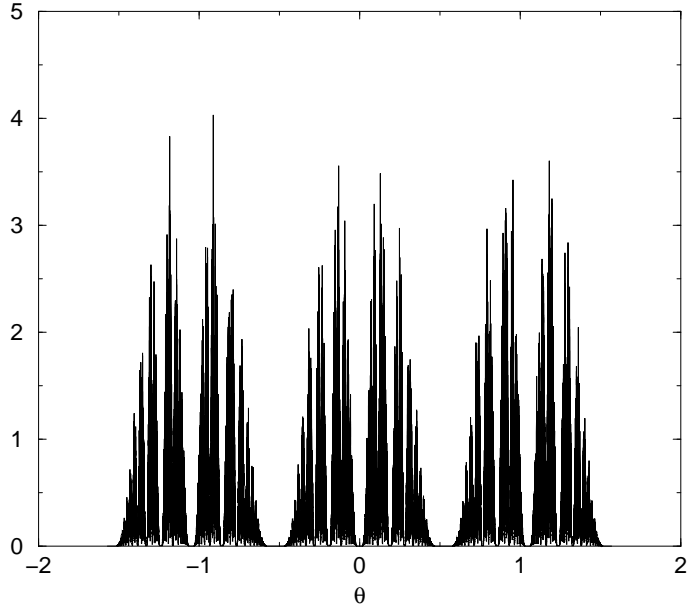,width=9cm}
\caption{Invariant measure $\overline\mu$ as a function of $\theta$}
\end{center}
\label{fig:5}
\end{figure}
The convergence $\mu^{(n)}(x)\to\overline{\mu}(x)$ for $n\to\infty$ is assured
by ergodic properties of such functional transform in case of equation
(\ref{rec2}), and has been successfully checked numerically by comparing to
direct sampling of different orbits. Obtaining $\mu_{\alpha}$ for
$\alpha=\{0,1,2\}$ from $\overline{\mu}$ is not difficult, taking into account
the symmetric role that they play with respect to the three intervals
$I_0=]-\infty,-1/\sqrt{3}],\ I_2=[-1/\sqrt{3},1/\sqrt{3}],\
I_1=[1/\sqrt{3},+\infty[$ (see fig.\ref{fig:5}).  Contracting properties
of $h_{\alpha}(x)$ allow convergence of (\ref{rec2}) only if
\be
\mu_k(x)=3\chi_{I_k}(x)\overline\mu(x)
\ee
where $\chi_{I_k}$ is the characteristic function of the interval $I_k$.

We would like to point out an interesting fact, even if far from being
rigorous, which is very similar to the argument put forward in \cite{gut} for
justifying the connexion between the invariant measure and the arithmetic
function $\beta(\xi)$. It has been shown in \cite{voitnech} that the lattice
under consideration can be isometrically embedded in a 2--manifold ${\cal
M}=\{c|\eta(z)|^4,\,z\in{\cal H}^2\}$, where
$$
\eta(z)=e^{\pi i z/12}\prod_{n=1}^{\infty}(1-e^{2\pi i n z})
$$
is the Dedekind $\eta$--function. The mountain range (relief) ${\cal M}$ displays a very steep
valley structure, and our tree lattice was defined as the ridges of this
relief. The natural ``counting'' of vertices whose real part lies in $[0,\xi]$
in \cite{gut} is in our case equivalent to counting the number of maxima of
$|\eta(x+i0^+)|^4$, that can be directly reexpressed as a density if one admits
that all maxima are equivalent and well separated:
\be \label{aut}
\overline\mu(x)\sim\frac{|\eta(x+i0^+)|^4}{\int_{0}^{1}| \eta(t+i0^+)|^4dt}
\ee
The intriguing fact is that $\eta^4$ is an automorphic form of weight 2, what
makes $|\eta|^4$ precisely a possible fixed point of Eq.(\ref{rec2}). We recall
the fundamental property of automorphic forms $f$ of weight 2 under the action
of $SL(2,\R)$:
\be
f(z)=\frac{e^{i\phi(a,b,c,d)}}{(cz+d)^2}f\left(\frac{az+b}{cz+d}\right)
\ee
The main problem is that the boundary behavior of automorphic forms is far from
trivial (see \cite{shein}), and (\ref{aut}) has no rigorous mathematical
sense.  In particular compatibility of (\ref{rec2}) and (\ref{aut}) is not
obvious even numerically. Nevertheless we insist on the fact that $\mu_k$ is
defined  with no ambiguity by (\ref{rec2}), what enables us to compute the
desired $W_n(d)$. The crucial point here, already required for convergence of
$\mu^{(n)}$, is ergodicity property of $\theta_n$. It means that for $n\gg
1$, the distribution of $\theta_n$ is exactly given by $\overline\mu(\theta)$,
independently of $n$ and initial condition. Then, denoting by $d_{n}^{\alpha}$
the value $d_n$ obtained for a word ending with $h_{\alpha}$, one can
transform (\ref{dist}) in the following way:
\be
d_{n+1}=\frac{1}{3}\sum_{\alpha=0}^{2}\left(d_{n}^{\alpha}+
\ln\left[\frac{5}{3}+\frac{4}{3}\cos(2\theta_n+\varphi_l)\right]\right)
\ee
with the condition $l\not=\alpha$. Thus we obtain
\be
\left<e^{ikd_n}\right>=\frac{1}{6}\sum_{k=0}^{2}\sum_{j\not=k}
\left[\int_{-\pi/2}^{\pi/2}d\theta\mu_k(\theta)\left(\frac{5}{3}+
\frac{4}{3}\cos(2\theta+\varphi_j)\right)^{ik}\right]^n
\ee
which finally leads to
\be
W_n(d)=\frac{1}{2\pi}\int_{-\infty}^{\infty}dk\,
e^{-ikd}\left[\int_{0}^{\frac{\pi}{3}}d\theta\mu_1(\theta-\frac{\pi}{6})
\left(\frac{5}{3}+\frac{4}{3}\cos2\theta\right)^{ik}\right]^n
\ee
This form suggests that for $n$ large $W_n(d)$ satisfies a central limit
theorem. Indeed such a theorem exists (see \cite{doob,boug}) for Markovian
processes, provided that the phase space is ergodic. We are then led to
compute only the first two moments (Lyapunov exponents) which gives
\be \label{gamma}
\gamma=\frac{\left<d\right>}{n}=
\int_{0}^{\frac{\pi}{3}}d\theta\mu_1(\theta-\frac{\pi}{6})
\ln\left(\frac{5}{3}+\frac{4}{3}\cos2\theta\right)\approx 0.792
\ee
and
\be
\sigma^2=\frac{\left<(d-\left<d\right>)^2\right>}{n}=\gamma_2-\gamma^2
\ee
with
\be \label{gamma2}
\gamma_2=\int_{0}^{\frac{\pi}{3}}d\theta\mu_1(\theta-\frac{\pi}{6})\ln^2
\left(\frac{5}{3}+\frac{4}{3}\cos2\theta\right)\approx 0.68
\ee
Numerical simulations presented in previous  section yield $\gamma\approx
0.793$ and  $\gamma_2\approx 0.66$, which finally allow us to conclude that
for $n\gg 1$ $W_n(d)$ has a Gaussian behavior
\be \label{gau}
W_n(d)=A\,e^{-\frac{(d-n\gamma)^2}{2\sigma^2n}}
\ee
centered at $\gamma n$ and of variance $\sigma^2n$ ($A$ is the
normalization).

The numerical values of the Lyapunov exponents $\gamma$ and $\gamma_2$ (see
Eqs.(\ref{gamma}) and (\ref{gamma2})) are obtained by means of semi--numerical
procedure which involves the numerical information about the invariant measure
$\mu_1(\theta)$. However one can get the estimates for the Lyapunov exponents
$\gamma$ and $\gamma_2$ by approximating the measure $\mu_1(\theta)$ on the
interval $0\le \theta\le \frac{\pi}{3}$ in two different ways:
\be \label{approx_mu}
\begin{array}{l}
\disp \mu_1(\theta-\frac{\pi}{6})\approx \mu_1^A(\theta)=\frac{3}{\pi}
\medskip \\
\disp \mu_1(\theta-\frac{\pi}{6})\approx \mu_1^B(\theta)=\frac{3}{2}
\sin(3\theta)
\end{array}
\ee
Both measures $\mu_1^A$ and $\mu_1^B$ are properly normalized on the interval
$[0,\frac{\pi}{3}]$. Substituting (\ref{approx_mu}) in (\ref{gamma}) and
(\ref{gamma2}) and computing (analytically for $\gamma$) the corresponding
integrals, one finally gets:
\be \label{approx_ga}
\mu_1^A:\ \left\{\begin{array}{l}
\gamma\approx 0.749 \\ \gamma_2\approx 0.665
\end{array} \right. \qquad
\mu_1^B:\ \left\{\begin{array}{l}
\gamma\approx 0.792 \\ \gamma_2\approx 0.684
\end{array}\right.
\ee
As one can see, the agreement between numerical values of Lyapunov exponents
obtained for the measures $\mu_1$ and its approximants $\mu_1^{A,B}$ is
reasonable for $\mu_1^A$ and very good for $\mu_1^B$.

\subsubsection{Multifractal exponents}

The partition function ${\cal Z}(q,N)$ introduced in (\ref{eq:sum}) can be
defined for any discrete subgroup of $PSL(2,\R)$ of generic element $\tau$ by
\be \label{eq:sum1}
{\cal Z}(q,N)=\sum_{\tau,\,l(\tau)\le N}e^{-qd(\tau)}
\label{eq:2}
\ee
and the associated critical exponent is then
\be
\tau(q)=-\lim_{N\to\infty}\frac{\ln {\cal Z}(q,N)}{N}
\ee

The probability distribution (\ref{gau}) enables us to rewrite (\ref{eq:sum1})
for the group $\Gamma_{\theta}$ in the limit $n\gg 1$ as follows
\be \label{s3}
{\cal Z}(q,N)=\sum_{n=1}^{N}3\times 2^{n-1}a_n
\ee
where
\be \label{s4}
a_n=\int_{0}^{\infty}e^{-qt}W_n(t)dt=
\int_{0}^{\infty}
%{\sqrt{2\pi n\sigma^2}}
A\,e^{-\frac{(t-n\gamma)^2}{2n\sigma^2}}e^{-qt}dt
\ee
The following two cases should be distinguished:
\begin{itemize}
\item For $q<\gamma/\sigma^2$, the minimum of the exponent in Eq.(\ref{s4}) is
within the range of integration and
\be
a_n\sim e^{-n\gamma q+n\sigma^2q^2/2}
\ee
hence
\be \label{s5}
\disp {\cal Z}(q,N)\sim \frac{3}{2}\sum_{k=1}^{N}e^{k(\ln 2-\gamma q+\sigma^2
q^2/2)}
\ee
The convergence of the sum (\ref{s5}) for $N\to\infty$ depends on the sign of
the function in the exponent. For
\be \label{eq:q0}
q<q_0=\frac{\gamma-\sqrt{\gamma^2-2\sigma^2\ln2}}{\sigma^2}
\ee
one has  $-\ln2+\gamma q-\sigma^2q^2/2>0$ and the multifractal exponent is
\be
\tau(q)=-\ln2+\gamma q-\sigma^2q^2/2
\ee
while for $q>q_0$, the series ${\cal Z}(q,N)$ 
converges and $\tau(q)=0$ what signals the termination of the
multifractality. Note that $q_0$ is real at least in the case of the group
$\Gamma_{\theta}$.

\item For $q>\gamma/\sigma^2$ the minimum of the exponent in Eq.(\ref{s4}) is
out of the range of integration and
\be
\ln a_n\sim-\frac{n\gamma^2}{2\sigma^2}
\ee
hence ${\cal Z}(q,N)$ is no longer extensive in $N$, which leads to $\tau(q)=
\tau(q_0)=0$.
\end{itemize}
The nonlinear dependence on $q$ obtained above shows the multifractal behaviour
of this model below the termination point $q_0$. It seems more transparent to
summarize all these results in a table

\begin{center}
\begin{tabular}{|c|c|} \hline
$q<\gamma/\sigma^2$  &
\begin{tabular}{ll}
$q<q_0=\frac{\gamma-\sqrt{\gamma^2-2\sigma^2\ln2}}{\sigma^2}$ &
$\qquad \tau(q)=-\ln2+\gamma q-\sigma^2q^2/2$ \\ \hline
$q>q_0=\frac{\gamma-\sqrt{\gamma^2-2\sigma^2\ln2}}{\sigma^2}$ &
$\qquad \tau(q)=0$ \\
\end{tabular} \\ \hline \hline
$q>\gamma/\sigma^2$ & $\tau(q)=\tau(q_0)=0$ \\ \hline
\end{tabular}
\end{center}

\subsubsection{Conformal mapping approach to computation of partition function
and multifractal exponent}

We propose in this section a completely different approach allowing to get a
closed analytic expression for the partition function similar to ${\cal
Z}(q,N_{\rm max})$ (see Eq.(\ref{eq:sum})). The construction presented below is
a by-product of our former investigations of analytic structure of the covering
Riemann space of the multi--punctured complex plane (see, for review
\cite{nech_rev}). We explore the properties of the Jacobian of conformal
mapping of the infinite complex plane with a triangular lattice of punctures
into the unit disc parametrized by $w=re^{i\alpha}$, which in this particular case represents the multi--sheeted
universal covering space \cite{nech_rev}. Namely, we define two functions
$f(r,\alpha)$ and $g(r)$:
\be \label{eq:fg}
f(r,\alpha)=c\frac{\left|\theta_1'\left(0,e^{i\pi \zeta(w)}\right)
\right|^{8/3}}{|1+iw|^4} \equiv
c\frac{\left|\eta(\zeta(w))\right|^8}{|1+iw|^4},
\quad g(r)=\frac{1}{(1-r^2)^2}
\ee
where
\be \label{eq:zeta}
\left\{\begin{array}{l}
\disp \theta_1'\left(0,e^{i\pi \zeta}\right)=
2\,e^{i\frac{\pi}{4}\zeta}\sum_{n=0}^{\infty}(-1)^n(2n+1)\,
e^{i\pi n(n+1)\zeta} \medskip \\
\disp \zeta(w)=e^{-i\pi/3}\,\frac{w+e^{i\pi/6}}{w-i} \medskip \\
\disp c=\left|\theta_1'\left(0,e^{i\pi \zeta(0)}\right) \right|^{-8/3}
\approx 0.933293
\end{array}\right.
\ee

\begin{figure}[ht]
\begin{center}
\epsfig{file=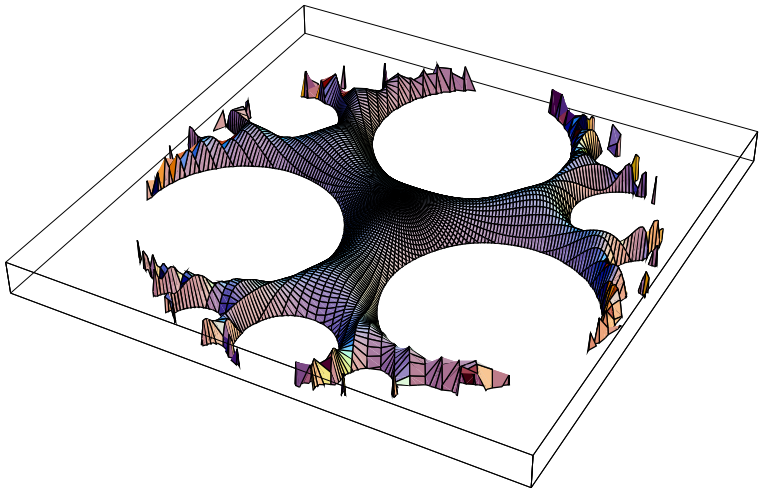,width=10cm}
\end{center}
\caption{3D parametric plot of the function $u(r,\alpha)$ in the
section $0.9<u(r,\alpha)\le 1$.}
\label{fig:6}
\end{figure}

One can show that the functional equation
\be \label{ea:vertices}
\frac{f(r,\alpha)}{g(r)}-1=0
\ee
has a family of solutions $(r_{\rm c},\alpha_{\rm c})$ exactly at positions
of 3--branching Cayley tree isometrically embedded in the hyperbolic unit 
disc (in the Klein's model of the surface of constant negative curvature).
In fig.\ref{fig:6} we have plotted the 3D section of the function
\be \label{eq:u}
u(r,\alpha)=\frac{f(r,\alpha)}{g(r)}
\ee
in polar coordinates $(r,\alpha)$ for $0.9<u(r,\alpha)\le 1$. The
function $u(r,\alpha)$ has local maxima {\it with one and the same value
$u=1$ only at the coordinates of isometric embedding of 3--branching Cayley
tree in the hyperbolic unit disc}. The proof of this fact is given in
Appendix \ref{app1}.

Thus, we can rewrite (\ref{eq:sum}) in the following closed form (recall that
Euclidean distance $r$ and the corresponding hyperbolic distance $d$ are linked
by the relation (\ref{eq:hyp}))
\be \label{eq:sum2}
\tilde{{\cal Z}}(q, d)=\frac{1}{2\pi}
\int\limits_{0}^{r(d)}\int\limits_{0}^{2\pi}\;
e^{-q\ln \frac{1+r}{1-r}}\delta\left(\ln
u(r,\alpha)\right)\left|\frac{d\ln u(r,\alpha)}{dr}\right|rdrd\alpha
\ee
where for $\delta[\ln u(x)]$  we use the standard integral representation
$\delta[\ln u(x)]=\frac{1}{2\pi}\int\limits_{-\infty}^{\infty}d\xi\,
[u(x)]^{i\xi}$.

It is noteworthy to pay attention to the difference between the partition
functions ${\cal Z}(q,N)$ (Eq.(\ref{eq:sum})) and $\tilde{{\cal Z}}(q, d)$
(Eqs.(\ref{eq:sum1}) and (\ref{eq:sum2})). The function ${\cal Z}(q,N)$ counts
the weighted number of Cayley tree vertices {\it up to the generation $N$} for
nonfixed maximal radius $r(d)=\tanh(d/2)$ in the hyperbolic unit disc, while 
the function $\tilde{{\cal Z}}(q, d)$ counts the weighted number of Cayley tree
vertices {\it within the hyperbolic disc of radius $r(d)$} for nonfixed maximal
generation $N$. The last partition function is in fact related to the number of
tree vertices inside the disc of radius $d$. This is the content of the famous
circle problem first formulated by Gauss for the Euclidean lattice $\Z^2$. The
extension to the non-Euclidean case is due to Delsarte \cite{delsarte} (see
also \cite{rudnick}).
% By conjecture the functions
%${\cal Z}(q,N)$ and $\tilde{{\cal Z}}(q, d)$ are
%related via the Legendre transform.

\section{Multifractality in 2D quasi--classical Liouville field theory}
\label{sect3}

Our starting point is the family of normalized wave functions $\psi_k({\bf x})$
defined as follows
\be \label{eq:gr}
\psi_k({\bf x})=\frac{|{\bf x}|^{k}e^{-\varphi({\bf x})}}{\left[\int
d{\bf x}\,|{\bf x}|^{2k}e^{-2\varphi({\bf x})}\right]^{1/2}}
\ee
where integration extends to a disc of radius $R$ and the potential
$\varphi({\bf x})$ is distributed with Gaussian distribution function
\be \label{eq:gr1}
P[\varphi({\bf x})]\propto \exp\left\{-\frac{1}{2g}
\int d{\bf x} \left(\nabla \varphi({\bf x})\right)^2\right\}
\ee
The problem defined in (\ref{eq:gr})--(\ref{eq:gr1}) appears in various
models which will be discussed in the next section.

The multifractal exponent for the quenched and annealed distributions
of disorder in (\ref{eq:gr})--(\ref{eq:gr1}) can be computed in the
standard way
\be \label{eq:quan}
\begin{array}{ll}
\disp \tau_{\rm qu}(q)=-\lim_{R\to \infty}\frac{\left<\ln Q(q,R)\right>}
{\ln R} & \mbox{for quenched disorder} \medskip \\
\disp \tau_{\rm an}(q)=-\lim_{R\to \infty}\frac{\ln\left<Q(q,R)\right>}
{\ln R} & \mbox{for annealed disorder}
\end{array}
\ee
where $Q(q,R)=\int d{\bf x}|\psi_{k}({\bf x})|^{2q}$ and the brackets
$\left<...\right>$ denote averaging with the distribution
$P[\varphi({\bf x})]$.

We pay attention to the case of annealed disorder and our aim is to evaluate
the correlation function
\be \label{eq:cor}
\left<Q(q,R)\right>=\int d{\bf x} \left<|\psi_{k}({\bf x})|^{2q}\right>
\ee
The averaging $\left<...\right>$ in (\ref{eq:cor}) means
\be \label{eq:aver1}
\left<...\right>_{S[\varphi]}=\int{\cal D}[\varphi({\bf x})]
e^{-S[\varphi({\bf x})]}
\ee
where
\be \label{eq:aver2}
S[\varphi({\bf x})]=\frac{1}{2g}\int d{\bf x}
\left\{\left(\nabla\varphi({\bf x})\right)^2\right\}
\ee

In order to take into account proper normalization of the wave function 
$\psi_{k}({\bf x})$ it is convenient to use a Lagrange multiplier $\lambda$, 
so that eventually
\be \label{eq:exp1}
\left<Q(q,R)\right>=\int d{\bf x}_0
\int{\cal D}[\varphi({\bf x})]e^{-2q\left(\varphi({\bf x}_0)-k\ln
|{\bf x}_0|\right)}\,e^{-S_{k}[\varphi({\bf x})]}
\ee
where
\be \label{eq:exp2}
S_{k}[\varphi({\bf x})]=\frac{1}{2g}\int d{\bf x}
\left\{\left(\nabla\varphi({\bf x})\right)^2+
2\lambda g\left(|{\bf x}|^{2k}e^{-2\varphi({\bf x})}-
\frac{1}{\pi R^2}\right)\right\}
\ee
is the action of 2D Liouville Field Theory (LFT).

The careful treatment of the quantum LFT in the case $k=0$ (see for review
\cite{gm}) enables one to find the conformal weights
$\Delta\left(e^{-2q\varphi}\right)= q({\cal Q}-q)$, where ${\cal Q}(g)$ is the
``background charge'', obtained by imposing conformal invariance of
$S_{0}[\varphi]$. The authors of work \cite{kogan} have related the value
$\Delta\left(e^{-2q\varphi}\right)$ to the critical exponent $\tau_{\rm an}(q)$
in the scaling dependence of the average inverse participation ratio
(\ref{eq:cor})
\be \label{eq:mult}
\left<Q(q,R)\right>\sim R^{-\tau_{\rm an}(q)}
\ee

Despite the multiscaling exponent $\tau_{\rm an}(q)$ has been computed in the
general framework of  Conformal Field Theory (CFT) few years ago, from our
point of view, the geometrical interpretation of the multifractal behavior in
the model has not yet been cleared up. A more physical approach put forward in
\cite{chamon} exploits an analogy between this model and the problem of
directed polymers on a Cayley tree. This analogy is supported by the fact that
in both cases the correlation functions grow logarithmically with the
distance. For directed polymers it is the correlation function of the random
potential defined on the tree vertices that scales logarithmically with the
ultrametric distance (i.e. distance along the tree). The same logarithmic
behaviour occurs in $2D$ Gaussian Field Theory \cite{kogan}.

We adopt a different point of view. Let us notice that the tree structure
(conjectured by C.Chamon et al) emerges quite naturally from the Liouville
field theory treated at a semi-classical level. Our idea is as follows. Indeed
there is not just one saddle point solution but a whole orbit of solutions
parametrized by $SL(2,\R)$. If one further assumes that the integration has to
be performed not over the whole group but only over a subgroup (for instance
$\Gamma_{\theta}$), one recovers quite naturally the model defined in section
\ref{sect2}.

Our starting point is the semi--classical ($g\rightarrow 0$) treatment of
(\ref{eq:exp1}) (a similar approach can be found in \cite{zam}). Using a
saddle point method, one is  led to the classical equation
\be \label{eq:euler}
\Delta\varphi -2qg\delta^2({\bf x}-{\bf x}_0)+
2\lambda g |{\bf x}|^{2k}\;e^{-2\varphi}=0
\ee
which gives, after integration
\be
\lambda=q-F/2g
\ee
where
$$
F=\int d{\bf x}\Delta\varphi
$$
is the magnetic flux. We then introduce the shifted field
\be
\tilde{\varphi}({\bf x})=\varphi({\bf x})-\ln|{\bf x}|^k
\ee
and taking into account that
\be
\Delta\tilde{\varphi}({\bf x})=\Delta\varphi({\bf x})-2\pi k\delta^2({\bf x})
\ee
we end up with the following equation
\be\label{col}
\Delta\tilde{\varphi}+ 2\pi
k\delta^2({\bf x})-2q\,g\delta^2({\bf x}-{\bf x}_0)+
(2q\,g-F)e^{-2\tilde{\varphi}}=0
\ee

The most general solution of (\ref{col}) (away from singularities) in Euclidean
space of complex coordinate $z$ can be written as follows \cite{gm}
\be \label{eq:class1}
e^{-2\tilde{\varphi}}=\frac{4}{|F-2qg|}\;\frac{\partial_z
A(z)\partial_{\overline{z}}B(\overline{z})}
{\Big(1+\epsilon A(z)B(\overline{z})\Big)^2}
\ee
where $A(z)$ and $B(\overline{z})$ are correspondingly holomorphic and
anti--holomorphic functions of $z$ and $\epsilon={\rm sign}(F-2q\,g)$.
%; $A(z)$ and $B(\overline{z})$ transform
%under the action of the groups $PSL(2,\R)$ or $PSU(1,1)$ depending on the
%choice of uniformization of the target space ${\cal H}^2$.
The semi--classical treatment assumes $g$ to be small, hence
in order to have real $\tilde{\varphi}$ in
Eq.(\ref{eq:class1}) we should put $\epsilon=1$, i.e. $q\ll F/2g$. The
relevant solution of (\ref{eq:class1}), compatible with the singularities,
reads
\be\label{sph}
e^{-2\tilde{\varphi}_{cl}(z)}=\frac{4(k+1)^2}{F}\frac{(z\bar
z)^k}{\left(1+(z\bar z)^{k+1}\right)^2}
\ee
The normalization condition of the wave function is then satisfied only if
$F=4\pi(k+1)$. We would like to stress that for $k=0$ we here recover the
critical value $4\pi$ of the magnetic flux: uniqueness of the ground state wave
function holds only below this value. It also should be mentioned that our
analysis does not depend on a peculiar basis of eigenfunctions and the
results presented here can be extended to wave functions of the form
$\psi(z)=P_k(z)e^{-\varphi(z)}$, $P_k$ being a polynomial of degree $k$. It is
noteworthy that (\ref{sph}) is an algebraically decaying wave function, what
is, following \cite{kogan}, a signature of the existence of prelocalized
states.

Using the fact that the Liouville field is not exactly a scalar but varies
under holomorphic coordinate transformations $z\rightarrow w(z)$ as
\be
\tilde{\varphi}(z)\to\tilde{\varphi}\left(w(z)\right)-\ln|w'(z)|,
\ee
one can check that the set of solutions (\ref{sph}) is invariant under the
following family of transformations, parametrized by the group $PSL(2,\C)$:
\be
z\to w_k(z;a,b,c,d)=\left(\frac{az^{k+1}+b}{cz^{k+1}+d}\right)^{\frac{1}{k+1}}
\ee
The orbit of $\tilde{\varphi}_{cl}(z)$ is then given by
\be
\label{orb}
\tilde{\varphi}_{cl}\left(z;a,b,c,d\right)=
\frac{1}{2}\ln\left(\frac{\pi\left(|az^{k+1}+b|^2+|cz^{k+1}+
d|^2\right)^2}{(z\bar z)^k}\right)
\ee
Up to redefinition of the measure $d\tau$ on $PSL(2,\C)$, we restrict the domain of integration to $PSL(2,\R)$.

Due to the angular symmetry of (\ref{sph}), we take points of the form
$z=i^{\frac{1}{k+1}}\rho$ ($\rho\in\R$), and following \cite{zam} we rewrite 
(\ref{eq:cor})--(\ref{eq:aver2})
\be\label{phicl}
\left<\rho^{2qk}e^{-2q\varphi(\rho)}\right>_{S_k[\varphi]}=
R^{-\frac{4\pi(k+1)^2}{g}}{\rm Det}\left[\frac{\delta^2S_k}{\delta\varphi^2}\right]^{-1/2}
\int_{PSL(2,\R)}e^{-2q\tilde{\varphi}_{cl}(\rho,\tau)}d\tau
\ee

%As discussed below, the integration over $PSL(2,\R)$ does not converge for all
% $q$: we therefore must introduce a
%cut-off $\Lambda$ such that $\int_{PSL(2,\R)}\equiv\int_{\ln{\rm Tr}(\tau\tau^
%{\dag)}\le
%\Lambda}\equiv\int_{\Lambda}$.
Let us denote
\be \label{eq:nu}
\tau=\left(\begin{array}{cc}
a & b \\ c & d
\end{array} \right)\quad {\rm and}\quad
\nu_{\rho}=\left(\begin{array}{ll}
\rho^{(k+1)/2} & 0 \\ 0 & \rho^{-(k+1)/2} \end{array}\right)
\ee
%is measure preserving on $PSL(2,\R)$ and roughly sets $\Lambda'\sim
%r^{k+1}\Lambda$,
then we can rewrite (\ref{phicl}) as follows
\be\label{phicl2}
\left<\rho^{2qk}e^{-2q\varphi(\rho)}\right>_{S_k[\varphi]}\propto
\rho^{-2q}\int_{PSL(2,\R)}e^{-2q\ln {\rm
Tr}\left[(\tau\nu_{\rho})(\tau\nu_{\rho})^{\dag}\right]}d\tau=
\rho^{-2q}I(\rho,q)
\ee
where we have got rid of irrelevant factors and the function $I(\rho,q)$ reads
\be\label{chvar}
I(\rho,q)=\int_{PSL(2,\R)}\left[2\cosh
d(i,\tau\nu_{\rho}(i))\right]^{-2q}d\tau
\ee

Instead of summing over the whole group $PSL(2,\R)$, we  restrict the sum over
a discrete subgroup, $\Gamma_{\theta}$ in our case. Even if this discretization
of the saddle manifold has no evident physical justification, we believe that
the model obtained yields interesting results. It leads to consider the
so-called Poincar\'e series (see \cite{els} for review) $H$, defined as follows
\be\label{poin}
I(\rho,q)=H(i,\nu_{\rho}(i),q)=
\sum_{\tau\in\Gamma_{\theta}}\left[2\cosh d(i,\tau\nu_{\rho}(i))\right]^{-2q}
\ee
As shown in \cite{els}, the series $H$ does not converge for  $q<q^{\star}$
with $q^{\star}$ depending on $\Gamma_{\theta}$ only (the analysis of the
previous sections show that we roughly may set $q^{\star}\simeq q_0/2$). A new dependence on $\rho$ occurs only if $H$ does not converge, we will therefore consider only this regime. We must introduce in this case a cut-off ${\cal N}$ to regularize the
series, and finally study asymptotics of the finite sum
\be\label{trunc}
I_{{\cal N}}(\rho,q)=H_{{\cal
N}}(i,\nu_{\rho}(i),q)=\sum_{\tau\in\Gamma_{\theta}/l(\tau)\le {\cal
N}}\left[2\cosh d(i,\tau\nu_{\rho}(i))\right]^{-2q}
\ee

This Poincar\'e series has the same asymptotic properties as the one that defines our model. In particular it will exhibit a multifractal behaviour in ${\cal N}$. However what  really matters for a physical system is the multifractal behaviour under transformations parametrized by $\rho$. We therefore have to relate the behaviour in the group manifold to the behaviour in the real space. This will be achieved through a renormalization transformation of the form

\be\label{rg}
I_{{\cal N}}(\rho,q)=C\rho^{\kappa}I_{{\cal N}'({\cal N},\rho)}(1,q)
\ee

Appendix \ref{app2} provides a heuristic derivation which gives
\be
I_{{\cal N}}(\rho,q)=C\rho^{-(k+1)\ln2/\gamma}I_{{\cal
N}+\frac{(k+1)}{\gamma}\ln\rho}(1,q)
\ee

Comparing (\ref{eq:sum1}) and (\ref{poin}) gives $I_{{\cal
N}}(1,q)\approx{\cal Z}(2q,{\cal N})$. Therefore
\be
I_{{\cal N}}(\rho,q)=C\rho^{-(k+1)\ln2/\gamma}{\cal Z}(2q,{\cal N}+\frac{k+1}{\gamma}\ln\rho)
\ee
This relation allows to extract the scale dependence in $\rho$ for a given cut-off ${\cal N}$. Using the asymptotics of ${\cal Z}$
obtained in previous sections yields
\be
\left<\rho^{2qk}e^{-2q\varphi(\rho)}\right>_{S_k[\varphi]}\propto\rho^{-2q-(k+1)\ln2/\gamma}\,\rho^{-\frac{k+1}{\gamma}\tau(2q)}
\ee
with the notations of previous sections. After integrating over the whole
domain we arrive at the final expression for the multifractal exponent
${\tau}_{\rm an}(q)$:
\be
\tau_{\rm an}(q)=-\lim_{R\to\infty}\frac{\ln\left<Q(q,R)\right>}{\ln R}=
\disp 2(q-1)\left(1-\frac{(k+1)\sigma^2}{\gamma}\,q\right)\ 
%\frac{1}{\gamma}(\tau(2q)-q\tau(2))
\mbox{for $q<q_0/2$}
\ee
with $q_0$ defined in (\ref{eq:q0}). The regular term $2(q-1)$ corresponds to the
one obtained in \cite{chamon} for $g\to 0$. Multifractality of the wave function
is induced by the quadratic term, which is directly related to geometric
properties of the saddle point hyperbolic manifold (target space), and holds
{\it in absence of any random potential} in this target space. The regime $q>q_0/2$ is not affected by these geometric properties.

\section{Conclusion}
\label{sect:4}

The wave function $\psi_{k}({\bf x})$ introduced in (\ref{eq:gr}) belongs
to the general class of exponential functionals of free fields. Such
functionals appear in several physical contexts.

1. The square of the wave function (for $k=0$) may be interpreted as the
equilibrium Gibbs measure in the random potential $\varphi({\bf x})$. In the 1D
case the problem was first studied in \cite{kree}. A rather deep and complete
analysis of this problem was recently presented in \cite{douss}.

2. Exponential functionals of free fields play an important role in the context of
one dimensional classical diffusion in a random environment. Their probability
distribution controls the anomalous diffusive behaviour of particles at large
time \cite{bouch}. They also arise in the study of disordered samples of finite
length \cite{oshan,monthus}. Some mathematical properties are discussed in 
\cite{yor}.

3. In the context of one dimensional localization in a random potential such
functionals arise in the study of the Wigner time delay \cite{tex2}.

4. The function $\psi_{k}({\bf x})$ is the ground state wave function of
2D Dirac fermions in a random magnetic field with $B=\Delta\varphi$. The
multifractal behaviour first conjectured in \cite{kogan} has been recently
confirmed by an independent investigation based on renormalization group
method \cite{douss}.  The scenario of multifractality which is presented here
%relies on a different approach since it
relies mainly on a geometric approach to a semiclassical quantization
scheme of the Liouville field theory. The fact that tree like structure
emerges quite naturally in our consideration  is an interesting feature
which obviously deserves further investigation. The multifractality in our
approach appears as a by-product of isometric embedding of a Cayley tree in
the hyperbolic plane. The objects which possess multifractal behavior are
the moments of the partition function defined as sums over all vertices of
a Cayley tree isometrically embedded in the hyperbolic plane where each
vertex carries a Botzmanm weight depending on the hyperbolic distance from
the root point. No randomness is imposed in the model.

From the mathematical side our work reveals some interesting links between the
theory of automorphic functions, invariant measures and spectral theory. We
hope to return to these problems in a forthcoming publication.

\noindent{\bf Acknowledgments}

The authors are grateful to Ph. Bougerol, J. Marklof, O. Martin, H. Saleur, and Ch. Texier
for valuable discussions and useful comments of different aspects of the problem.

\begin{appendix}
\section{}
\label{app1}

Let us prove that the function $u(w)=(1-w\overline{w})^2\;f(w)$
where $f(w)$ is defined in (\ref{eq:fg}) has the following properties:
\begin{itemize}
\item At all centers $w=w_c$ of zero--angled triangles tesselating the
Poincar\'e hyperbolic unit disc $u(w_c)=1$;
\item The function $u(w)$ has local maxima at the points $w_c$.
\end{itemize}
I. The proof of the first statement implies the proof of the fact that the function
$u(w)$ is invariant with respect to the conformal transform $w^{(1)}(w)$
of the unit Poincar\'e disc to itself where
\be \label{A.1}
w^{(1)}(w)=\frac{w-w_0}{w\overline{w}_0-1},
\ee
and $w_0$ is the coordinate of any center of zero--angled triangle in the
hyperbolic Poincar\'e disc obtained by successive transformations from the initial
one.

Hence, it is neccessary and sufficient to show that the values $u(w=0)$,
$u\left(w=-\frac{i}{2}\right)$, $u\left(w=\frac{1}{2}e^{i\pi/6}\right)$ and
$u\left(w=\frac{1}{2}e^{i 5\pi/6}\right)$ coincide. Then, performing the
conformal transform and taking $w_0=\left\{-\frac{i}{2},\;
\frac{1}{2}e^{i\pi/6},\;\frac{1}{2}e^{i 5\pi/6}\right\}$, we move the
centers of the first generation of zero--angled triangles to the center of
the disc $w^{(1)}$. Now we can repeat recursively the contruction, i.e.
find the new coordinates of the centers of the second generation of
zero--angled triangles in the disc $w^{(1)}$ and compute the function
$u(w)$ at these points, then we perform the conformal transform
$w^{(2)}(w^{(1)})$ and so on...

We have at the point $w=0$:
$$
u(w=0)=c\left|\theta_1'(0,e^{i \pi[1/2+i\sqrt{3}/2]})\right|^{8/3}=1
$$
while at the point $w=-\frac{i}{2}$ the function $u(w)$ can be written in
the form
\be \label{A.2}
u\left(w=-\frac{i}{2}\right)=
\frac{c}{\left|1-\frac{1}{2}e^{-i\pi}\right|^4}
\left|\theta_1'(0,e^{i\pi[1/2+i/(2\sqrt{3})]})\right|^{8/3}
\left(1-\frac{1}{4}\right)^2
\ee

Let us use the properties of Jacobi $\theta$--functions:
\be \label{A.3}
\left\{\begin{array}{ll}
\theta_1'(0,e^{i\pi(w+k)})=
\theta_1'(0,e^{i\pi w}); & \quad k\in N  \medskip \\
\theta_1'(0,e^{i\pi[1/2+i/(2\sqrt{\lambda})]})\lambda^{-3/4}=
\theta_1'(0,e^{i\pi[1/2+i\sqrt{\lambda}/2]}); & \quad \lambda\in R
\end{array}\right.
\ee
Taking into account (\ref{A.3}) we can rewrite (\ref{A.2}) in the form
\be \label{A.4'}
\begin{array}{lll}
\disp u\left(w=-\frac{i}{2}\right) & = &
\disp \frac{2^4\;c}{3^4}\;(3^{3/4})^{8/3}
\left|\theta_1'(0,e^{i\pi[1/2+i/(2\sqrt{3})]3})\right|^{8/3}
\left(\frac{3}{4}\right)^2 \medskip \\
& = & \disp c\left|\theta_1'(0,e^{i\pi(1/2+i\sqrt{3}/2)+i\pi})\right|^{8/3}=1
\end{array}
\ee
Thus, $u(w=0)=u\left(w=-\frac{i}{2}\right)=1$.

At the point $u\left(w=\frac{1}{2}e^{i\pi/6}\right)$ we have
\be \label{A.4''}
\begin{array}{lll}
\disp u\left(w=\frac{1}{2}e^{i\pi/6}\right) & = & \disp \frac{2^4\;c}{3^2}
\left|\theta_1'(0,e^{i\pi[3/2+i\sqrt{3}/2]})\right|^{8/3}
\left(1-\frac{1}{4}\right)^2 \medskip \\ & = &
c\left|\theta_1'(0,e^{i\pi[1/2+i\sqrt{3}/2]+i\pi})\right|^{8/3}=1
\end{array}
\ee
In the same way we can transform the function
$u\left(w=\frac{1}{2}e^{i 5\pi/6} \right)$:
\be \label{A.4'''}
\begin{array}{lll}
\disp u\left(w=\frac{1}{2}e^{i 5\pi/6}\right) & = & \disp\frac{2^4\;c}{3^2}
\left|\theta_1'(0,e^{i\pi[-1/2+i\sqrt{3}/2]})\right|^{8/3}
\left(1-\frac{1}{4}\right)^2 \medskip \\ & = &
c\left|\theta_1'(0,e^{i\pi[1/2+i\sqrt{3}/2]-i\pi})\right|^{8/3}=1
\end{array}
\ee
The transforms (\ref{A.4'})--(\ref{A.4'''}) complete the proof of the part I.

II. Let us prove that the function $u(w)$ has local maxima at all centers
$w_c$ of zero--angled triangles tesselating the Poicar\'e hyperbolic disc.
Actually, the function $f(w)$ by construction gives a metric of some
discrete subgroup of the group of motions of Poicar\'e hyperbolic disc.
Hence the function $f(w)$ cannot grow faster than the isortopic hyperbolic
metric $(1-w\overline{w})^2$ and the following inequality is valid
$$
0<u(w)\le 1
$$
for all points $w$ inside the unit disc. But we have shown that $u(w)=1$ at
$w=w_c$ what means that the function $u(w)$ reaches its local maxima at the
points $w_c$ and at all these maximal points the function $u(w)$ has one
and the same value $u(w_c)=1$. The part II is proved.

\section{}
\label{app2}

Our aim is to extract explicitly the $\rho$--dependence of the truncated series
(\ref{trunc}) and to connect it  to $I(1,q)$. More precisely we are looking for
a renormalization transformation of the form
\be\label{rg2}
I_{{\cal N}}(\rho,q)=C\rho^{\kappa}I_{{\cal N}'({\cal N},\rho)}(1,q)
\ee

\begin{figure}[ht]
\begin{center}
\epsfig{file=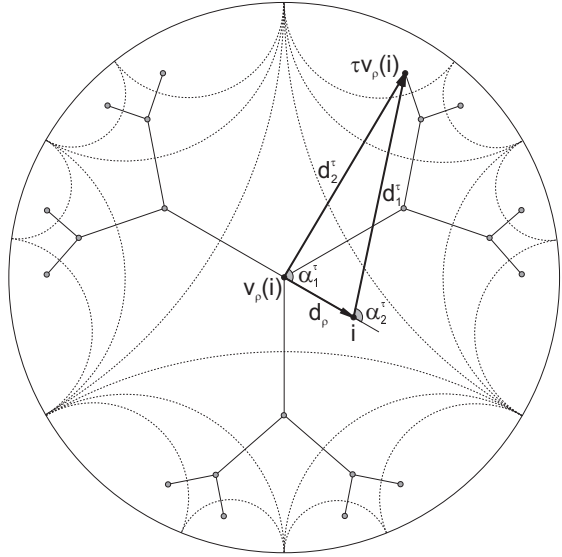,width=5cm}
\end{center}

\caption{Change of coordinates in hyperbolic Poincar\'e disc}
\label{fig:7}
\end{figure}

Using the correspondance (up to the volume of $SO(2)$) between $PSL(2,\R)$ and
${\cal H}^2$, we interpret the shift $\tau\to\tau\nu_{\rho}$ as a change of hyperbolic coordinates---see fig \ref{fig:7}.

Note that the expression (\ref{trunc}) does not depend on the particular
representation of the hyperbolic 2-space, since the hyperbolic distance is
invariant. The only one requirement is to define a compatible action of
$\Gamma_{\theta}$ in the space under consideration. We use for conveniency
the unit disc representation whose center is the image of the point
$\nu_{\rho}(i)$ where $\nu_{\rho}$ is defined by Eq.(\ref{eq:nu})
%(refered to as $\nu_{\rho}(i)$)
in the ${\cal H}^2$ representation. For shortness the generic element of
$\Gamma_{\theta}$ is labelled by $\tau$ independent on the representation. We
parametrize the point $\tau\nu_{\rho}(i)$ by its hyperbolic polar coordinates
$(d^{\tau}_1,\alpha^{\tau}_1)$ with the origin at the point $i$ and by
``shifted'' hyperbolic coordinates $(d^{\tau}_2,\alpha^{\tau}_2)$ with the
origin at the point $\nu_{\rho}(i)$. Note that $d_{\rho}=d(i,\nu_{\rho}(i))
\sim(k+1)\ln\rho$, and the following ``triangle equation'' in hyperbolic 2-space
holds:
\be
\cosh d^{\tau}_1=\cosh d^{\tau}_2\cosh d_{\rho}+
\sinh d^{\tau}_2\sinh d_{\rho}\cos\alpha^{\tau}_2
\ee
%Instead of summing over the whole group $PSL(2,\R)$, we  restrict the sum over
% a discrete subgroup, $\Gamma_{\theta}$ in our case.
In order to extract the scaling conjectured in (\ref{rg}), we make an
approximation which consists in neglecting fluctuations of $W_n(d)$. In this
approximation we can sum over the generations $n$ and the angles
$\alpha_{j_n}$ of the vertices within each generation ($1\le j_n\le
3\times 2^{n-1}$). Namely $(d^{\tau}_2,\alpha^{\tau}_2)=(\gamma n,
\alpha_{j_n})$. Thus one has
\be\label{disc}
I_{{\cal N}}(\rho,q)=\sum_{n=1}^{{\cal N}}\left[2\cosh\gamma n\cosh
d_{\rho}\right]^{-2q}\sum_{j_{n}=1}^{3\times 2^{n-1}}(1+\tanh \gamma n\tanh
d_{\rho}\cos\alpha_{j_{n}})^{-2q}
\ee
Assuming that $\alpha_{j_{n}}$ are uniformly distributed, we get for $n\gg 1$
the following expression
\be\label{eq:angles}
\sum_{j_{n}=1}^{3\times 2^{n-1}}(1+\tanh \gamma n\tanh
d_{\rho}\cos\alpha_{j_{n}})^{-2q}\approx \frac{3\times
2^{n-1}}{2\pi}\int_{0}^{2\pi}\frac{d\alpha}{(1+\tanh \gamma n\tanh
d_{\rho}\cos\alpha)^{2q}}
\ee
which leads to the asymptotic behavior:
\be\label{asym}
\left.\lim_{n\to\infty}2^{-n}\sum_{j_{n}=1}^{3\times 2^{n-1}}
(1+\tanh \gamma n\tanh d_{\rho}\cos\alpha_{j_{n}})^{-2q}
\right|_{\rho\to\infty}\sim
\left\{\begin{array}{ll} {\rm const} & \mbox{for $q\le 1/4$} \medskip \\
e^{2d_{\rho}(2q-\frac{1}{2})} & \mbox{for $q>1/4$} \end{array}\right.
\ee
 As justified hereafter we consider as relevant only the case $q\le1/4$. Using that for $n\gg 1$ and $\rho\gg 1$
one has $2\cosh\gamma n\cosh d_{\rho}\sim\cosh(\gamma n+d_{\rho})$ we can
rewrite Eq.(\ref{disc}) as follows
\be\label{disc2}
I_{{\cal N}}(\rho,q)= C\rho^{-(k+1)\ln2/\gamma}\sum_{n=1}^{{\cal
N}}2^{(n+d_{\rho}/\gamma)}[\cosh(\gamma n+d_{\rho})]^{-2q}
\ee
Performing the shift $\tilde{n}=n+d_{\rho}/\gamma$ we get finally
\be
I_{{\cal N}}(\rho,q)=C\rho^{-(k+1)\ln2/\gamma}I_{{\cal
N}+\frac{(k+1)}{\gamma}\ln\rho}(1,q)
\ee
This expression fulfills the condition (\ref{rg2}). We assume that this
renormalization also holds for the full function $I_{{\cal N}}(\rho,q)$.
%, at least at first order level.

Let us pay attention to some contradiction between (\ref{chvar}) and (\ref{asym})
raised by the set of successive approximations of (\ref{chvar}) which however is
irrelevant for our final conclusions about multifractality. The equation
(\ref{chvar}) shows that if the integral over $PSL(2,\R)$ converges, it
should not depend on $\rho$. Using the Poincar\'e series, the convergence of
(\ref{chvar}) occurs for $q>q^{\star}$. For $q>1/4$ and $q>q^{\star}$ the
$\rho$--dependence shown in (\ref{asym}) should then cancel by summing over all $n$. The discrepancy between (\ref{chvar})
and (\ref{asym}) appears for $q\in[1/4, q^{\star}]$. First of all we should note
that the interval $[\frac{1}{4}, q^{\star}]$ is numerically small (following
previous sections we have $q^{\star}\simeq q_0/2\simeq 0.4$) and is nonuniversal,
i.e. depends on the particular choose of the subgroup under consideration.
Moreover, both the threshold $q=1/4$ and the asymptotics (\ref{asym}) depend on
the distribution of $\alpha_{j_n}$ and we believe that more careful treatment of
angle dependence in (\ref{chvar}) would allow disregard the region
$[1/4, q^{\star}]$.
\end{appendix}

\end{document}